\documentclass{article}
\usepackage{hiph-art}
\usepackage{epsfig}
\usepackage{amssymb}
\newcommand{\beq}{\begin{equation}}
\newcommand{\eeq}{\end{equation}}
\newcommand{\bea}{\begin{eqnarray}}
\newcommand{\eea}{\end{eqnarray}}
\newcommand{\pa}{\parallel}
\newcommand{\pe}{\perp}
\newcommand{\bx}{{\vec x}}
\newcommand{\by}{{\vec y}}

\volnumber{} \issuenumber{} \edyear{}                             %|
\frompage{} \topage{}                                              %|
\recrevdate{1 December 2005}                                              %|
\def\rightmark{Nonequilibrium quantum fields}
\def\leftmark{J.\ Berges and Sz.\ Bors\'anyi}
 
\title{Nonequilibrium quantum fields from first principles} 
\authors{ 
{J\"urgen Berges and Szabolcs Bors\'anyi %
\index{Berges, J.} % Abbreviated names of the author(s),
\index{Bors\'anyi, Sz.} % to be inserted for use in the volume index
}\\[2.812mm]
{\normalsize
Universit\"at Heidelberg, Institut f\"ur
Theoretische Physik,\\ 
Philosophenweg 16, 69120 Heidelberg, Germany\\[0.2ex] 
}}
 
\abstract{Calculations of nonequilibrium processes become increasingly
feasable in quantum field theory from first principles. There has been 
important progress in our analytical understanding based on 2PI generating 
functionals. In addition, for the first time direct lattice simulations 
based on stochastic quantization techniques have been achieved. 
The quantitative descriptions of characteristic far-from-equilibrium 
time scales and thermal equilibration in quantum field theory
point out new phenomena such as prethermalization. They determine the 
range of validity of standard transport or semi-classical approaches, 
on which most of our ideas about nonequilibrium dynamics were based so far.
These are crucial ingredients to understand important
phenomena in high-energy physics related to collision 
experiments of heavy nuclei, early universe cosmology and complex
many-body systems. 
}

\keyword{nonequilibrium, 2PI effective action, lattice simulations}

\PACS{11.10.Wx,12.38.Mh,05.70.Ln}
 
\makeindex
\begin{document}
 
\maketitle

\section{Characteristic far-from-equilibrium time scales}\label{intro}

Understanding the dynamics of quantum fields far away from the ground 
state or thermal equilibrium is a challenge touching many aspects of physics,
ranging from early cosmology or collision experiments with heavy nuclei
to ultracold quantum gases in the laboratory.
One of the most crucial aspects concerns the characteristic time
scales on which thermal equilibrium is approached. 
Much of the recent interest 
derives from observations in collision experiments of heavy nuclei at RHIC.  
The experiments seem to indicate the early validity of hydrodynamics after
somewhat less than $1\,$fm, 
whereas the present theoretical understanding of 
QCD suggests a longer thermal equilibration time. 

\begin{figure}[t]
\begin{center}
\epsfig{file=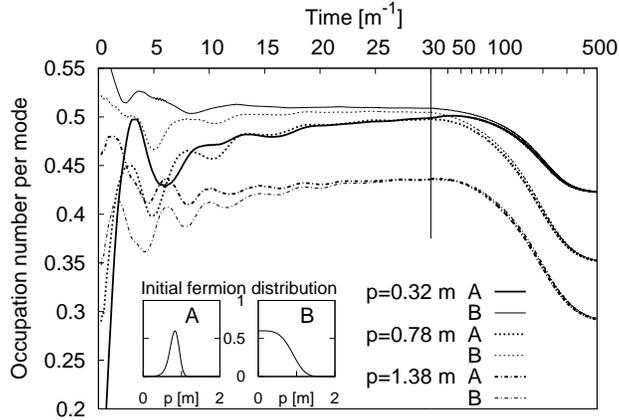,width=8.5cm}
\end{center}
\vspace*{-0.7cm}
\caption{Fermion occupation number 
for three different momentum modes as a function of time
in the chiral quark meson model of Ref.~\cite{Berges:2004ce}.
} 
\label{fig:join_fn}
\end{figure}

Here it is important to note that different quantities effectively thermalize 
on different time scales and a complete thermalization of all 
quantities may not be necessary to explain the observations.
This has been pointed out in Ref.~\cite{Berges:2004ce}, where it was shown 
for a chiral quark-meson model that the prethermalization of important 
observables occurs on time scales dramatically shorter than the thermal 
equilibration time. As an example, Fig.~\ref{fig:join_fn} shows 
the nonequilibrium time 
evolution of fermion occupation number for three different momentum modes
in this model. The evolution is given for two different initial
particle number distributions A and B shown in the insets, 
with {\em same} energy density. Therefore, both runs have to lead
to the same distributions in thermal equilibrium. The vertical line 
in Fig.~\ref{fig:join_fn} marks the characteristic time scale 
$\sim t_{\rm damp}$, after which the details about the initial 
distributions A or B are effectively lost. One observes that this
happens far before the late-time approach to thermal equilibrium.
The time $t_{\rm damp}$ for the effective loss of initial 
condition details is an important characteristic time scale
in nonequilibrium dynamics, which is very different from the
thermal equilibration time $t_{\rm eq}$. Note that 
the long-time behavior to thermal equilibrium is shown on a 
logarithmic scale in Fig.~\ref{fig:join_fn}
(in units of the scalar thermal mass $m$). 

\begin{figure}[t]
\begin{center}
\epsfig{file=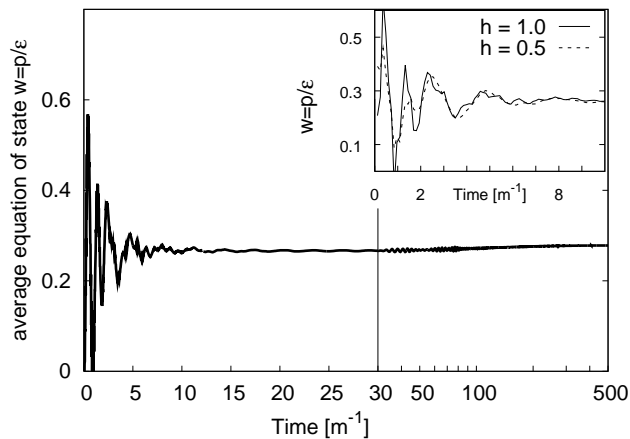,width=8.5cm}
\end{center}
\vspace*{-0.7cm}
\caption{The ratio of average pressure over energy density $w$ 
as a function of time. The inset shows the early stages 
for two different couplings $h$ and demonstrates
that the prethermalization
time is rather independent of the interaction details.}
\label{fig:wevol}
\end{figure}

In contrast to the very long time $t_{\rm eq}$ for 
complete thermal equilibration, prethermalization of the (average) 
equation of state sets in extremely rapidly on a time scale
\beq
t_{\rm pt} \ll t_{\rm damp} \ll t_{\rm eq} \, .
\eeq 
In Fig.~\ref{fig:wevol} 
we show the ratio of average pressure (trace over space-like components
of the energy-momentum tensor) over energy density,
$w = p/\epsilon$, as a function of time. One observes that
an almost time-independent equation of state builds up 
very early, even though the system is still far from equilibrium!
Here the prethermalization time $t_{\rm pt}$ is 
of the order of the characteristic
inverse mass scale $m^{-1}$. This is a typical consequence of
the loss of phase information by summing over oscillating functions 
with a sufficiently dense frequency spectrum.
If the ``temperature'' ($T$), i.e.~average kinetic energy per mode,
sets the relevant scale one finds  
$T\, t_{\rm pt} \simeq 2 - 2.5$~\cite{Berges:2004ce}.
For $T \gtrsim 400 - 500\,$MeV one
obtains a very short prethermalization
time $t_{\rm pt}$ of somewhat less than $1\,$fm.

This is consistent with very early hydrodynamic behavior, however,
it is not sufficient as noted in Refs.~\cite{Berges:2004ce,Arnold:2004ti}.
Beyond the average equation of state, a crucial ingredient
for the applicability of hydrodynamics for collision 
experiments~\cite{Heinz:2004pj} is the approximate isotropy 
of the local pressure. More precisely, the diagonal (space-like) components
of the local energy-momentum tensor have to be approximately equal.
Of particular importance is the possible isotropization far from 
equilibrium. The relevant time scale for the early validity 
of hydrodynamics could then be set by the isotropization 
time. The analysis of simple models lead to an isotropization
time which is given by the characteristic damping time 
$\sim t_{\rm damp}$ (cf.~Figs.~\ref{fig:join_fn})~\cite{Berges:2005ai}. 
However, a possible weak-coupling mechanism for faster isotropization 
in gauge theories such as QCD has been identified in 
Ref.~\cite{Arnold:2004ti} in terms of plasma 
instabilities~\cite{Mrowczynski:1993qm}. Whether this can explain
the experimental observations or whether they suggest that we have 
to deal with some new form of a ``strongly coupled Quark Gluon Plasma''
is an important open question. 

\section{What can we learn from transport or kinetic theory?}

For out-of-equilibrium calculations there are
additional complications, which do not appear in thermal
equilibrium or vacuum. Standard approximation techniques, such 
as perturbation theory, are not uniform in time and fail to 
describe thermalization. 
Aspects of systems with high occupation numbers are often 
successfully described using classical field theory methods.
However, classical Rayleigh-Jeans divergences and the lack of 
genuine quantum effects --- such as the approach to quantum 
thermal equilibrium characterized by Bose-Einstein or 
Fermi-Dirac statistics --- limit their use. 

Most theoretical approaches to the important question of
thermalization have been limited to semi-classical 
systems in the weak-coupling limit so far. With the advent of new
computational techniques a more direct account of quantum 
field degrees of freedom becomes more and more possible. 
There has been important progress in our 
understanding of nonequilibrium quantum fields using 
suitable resummation techniques based on 2PI generating
functionals~\cite{Berges:2004yj}. 
They have led to quantitative descriptions of 
far-from-equilibrium dynamics and thermalization in a variety 
of scalar and fermionic quantum field theories so far. An example for
the results of such a first-principles calculation within
a quark-meson model in 3+1 dimensions is given in 
Figs.~\ref{fig:join_fn} and \ref{fig:wevol}.
One important application of these quantum field theoretical 2PI methods 
is to test standard transport or semi-classical approaches, on which most
of our ideas about nonequilibrium dynamics are based so far. 

Several of these tests have been done for simple scalar 
$\lambda \Phi^4$ quantum field theories for not too strong 
couplings $\lambda$, which are well under quantitative control 
using 2PI techniques. 
Following Ref.~\cite{Berges:2005ai}, we consider a class of anisotropic
initial conditions with an initially high occupation number 
of modes moving in a narrow momentum range around the ``beam direction''
$p_3 \equiv p_\pa = \pm p_{\rm ts}$. The spatially homogeneous occupation
numbers for modes with momenta perpendicular to this direction, 
$p_1^2 + p_2^2 \equiv p_{\pe}^2$, are small or vanishing.
The situation is reminiscent of some aspects of the anisotropic
initial stage in the central region of two colliding wave 
packets.\footnote{Other interesting scenarios
include ``color-glass''-type initial conditions with 
distributions $\sim \exp(-\sqrt{p_\pe^2}/Q_s)$ peaked around $p_3 =0$
with ``saturation'' momentum~$Q_s$.} Of course,
a peaked initial particle number distribution is not
very specific and is thought to exhibit characteristic properties
of nonequilibrium dynamics for a large variety
of physical situations.
\begin{figure}[t]
\begin{center}
\epsfig{file=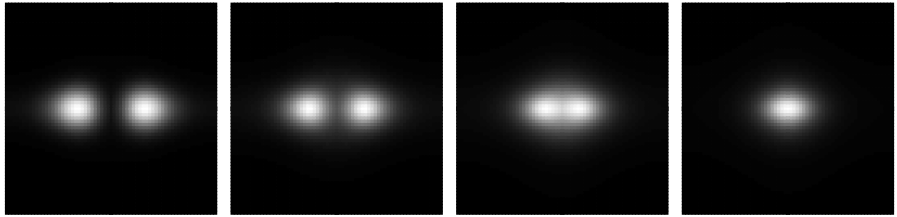,width=12.cm}
\end{center}
\vspace*{-0.7cm}
\caption{
Snapshots at times 
$t \simeq 0$, $t_{\rm damp}/2$, $t_{\rm damp}$ and $3t_{\rm damp}/2$ 
of the initially anisotropic occupation number distribution as 
a function of $p_{\pe}$ (vertical) and 
$p_{\pa}$ (horizontal). The shown resolution
was achieved by using a $64^3$ spatial lattice.} 
\label{fig:snapshots}
\end{figure}

An example of the earlier stages of such an evolution is shown 
in Fig.~\ref{fig:snapshots}.
Shown are snapshots of the occupation number distribution 
as a function of $p_{\pe}$ (vertical) and 
$p_{\pa}$ (horizontal), where bright (dark) regions correspond
to high (low) occupation numbers. The initial Gaussian distribution
is centered around a momentum of the order of the
renormalized thermal mass, and we will 
consider amplitudes of order one for the comparisons below 
(see \cite{Berges:2005ai}
for details). From Fig.~\ref{fig:snapshots} it can be seen that after the  
characteristic damping time $t_{\rm damp}$ the distribution starts to 
become rather independent of the momentum direction. 
Finally, at about $3t_{\rm damp}/2$ the figure shows an almost perfectly 
isotropic situation. We emphasize that the distribution is still
far from equilibrium. The situation is similar to what is displayed
in Fig.~\ref{fig:join_fn}, where the vertical line indicates about
$t_{\rm damp}$. In general, we find that the characteristic damping 
time is well described by the standard relaxation-time approximation
\beq
t_{\rm damp} \,\simeq\, 
- \frac{2 \omega^{(\rm eq)}}{\tilde{\Sigma}^{(\rm eq)}_{\varrho}}
\, \simeq\, \frac{4 \pi m}{3 \lambda^2 T^2} \, ,
\label{eq:estimate}
\eeq
where the imaginary part of the thermal equilibrium 
self-energy $-\tilde{\Sigma}^{(\rm eq)}_{\varrho}/2$ in Fourier-space
is evaluated for on-shell frequency $\omega^{(\rm eq)}$ for 
momentum $p_{\rm ts}$. The second equality in (\ref{eq:estimate})
is only valid for sufficiently high temperatures and
weak couplings. We emphasize that the relaxation-time approximation  
(\ref{eq:estimate}) does not describe the thermalization time 
$t_{\rm eq}$~\cite{Berges:2005ai}.

The question is whether transport or kinetic equations can be used 
to quantitatively describe the early-time behavior $t < t_{\rm damp}$, 
which is necessary if their application to the problem of fast 
apparent thermalization explained in Sec.~\ref{intro} is viable. 
The derivation of transport equations is standard. 
The evolution of particle number distributions is encoded in  
the time-ordered two-point correlation function.
Its imaginary and real part is determined by the commutator and
anti-commutator of two fields:
\beq
\rho(x,y) = i \langle [\Phi(x),\Phi(y)] \rangle \, , \quad
F(x,y) = \frac{1}{2} \langle \{\Phi(x),\Phi(y)\} \rangle \, .
\label{eq:anticomF}
\eeq
Here $\rho(x,y)$ denotes the spectral function and 
$F(x,y)$ the statistical two-point function.
While the spectral function encodes the
spectrum of the theory, the statistical 
propagator gives information about occupation numbers.
Loosely speaking, the decomposition makes explicit
what states are available and how they are occupied.  
For nonequilibrium
$F(x,y)$ and $\rho(x,y)$ are in general two independent 
two-point functions, whose exact time evolution equation
reads\footnote{For a detailed derivation see e.g.\ Ref.~\cite{Berges:2004yj}.
We are considering the equations for Gaussian 
initial conditions, which are underlying transport equations.
More involved initial conditions can be considered using
higher $n$-PI effective actions~\cite{Berges:2004yj}.} 
\bea 
\left[ \square_x 
+ M^2(x) \right] F(x,y) &=& 
- \int_0^{x^0}\! {\rm d}z^0 
\int_{-\infty}^{\infty}\! {\rm d}^3 z\,
\Sigma_{\rho}(x,z) F(z,y)
\nonumber\\
&& + \int_0^{y^0}\! {\rm d}z^0 
\int_{-\infty}^{\infty}\! {\rm d}^3 z\, 
\Sigma_F(x,z) \rho(z,y)  \, ,
 \label{eq:exactevolF} \\[0.1cm]
\left[\square_x + M^2(x) 
\right] \rho(x,y) &=& 
- \int_{y^0}^{x^0}\! {\rm d}z^0 
\int_{-\infty}^{\infty}\! {\rm d}^3 z\, 
\Sigma_{\rho}(x,z) \rho(z,y) \, .
\label{eq:exactevolrho}
\eea
These are causal equations with characteristic 
``memory'' integrals, which integrate over the time history of the
evolution starting at time $t_0=0$. Since they are exact they are equivalent 
to any kind of identity for the two-point functions such as 
Schwinger-Dyson/Kadanoff-Baym equations. Here
$\Sigma_F(x,y)$ denotes the real part and $-\Sigma_\rho(x,y)/2$ 
the imaginary part of the self-energy $\Sigma$, where the local
contribution is taken into account in $M^2(x)$. In 2PI
approximations the self-energy $\Sigma$ is obtained from 
the two-particle irreducible effective action~\cite{Cornwall:1974vz}.
Here we consider a three-loop 2PI effective action~\cite{Berges:2005ai}. 
It includes direct scatterings, off-shell and memory effects. Most
importantly in this context, it employs no derivative expansion.
The latter is a basic ingredient for transport or kinetic theory.

Transport equations are obtained from the exact equations by the
following prescription: 

1) The lower bound ($t_0=0$) of the time-integrals in
(\ref{eq:exactevolF}) is sent to the infinite past, i.e.\ $t_0 \to -\infty$. 
Of course, a system that thermalizes would have reached already
equilibrium at any finite time if initialized in the remote past. 
Therefore, in practice a ``hybrid'' description is employed: 
transport equations are
initialized by prescribing $F$, $\rho$ and derivatives 
at a {\em finite} time using the equations with $t_0 \to -\infty$ 
as an approximate description. 

2) Employ a gradient expansion. In practice, this expansion is carried 
out to lowest order (LO) or 
next-to-lowest order (NLO) in the number of derivatives with respect to
the center coordinates $X^{\mu} \equiv (x^{\mu} + y^{\mu})/2$ 
and powers of the relative coordinates
$s^{\mu} \equiv x^{\mu} - y^{\mu}$.

3) Even for finite $X^0$ one assumes that the 
relative-time coordinate $s^0$ ranges from $-\infty$ to $\infty$
in order to achieve a convenient description in Wigner space, i.e.\ in Fourier 
space with respect to the relative coordinates.  

We emphasize that the {\em ad hoc} approximations 1) and 3)
are in general not controlled by a small expansion parameter.
They require a loss of information about the details of the initial state.
More precisely, they can only be expected to be valid for sufficiently
late times times $t$ when initial-time correlations   
become negligible, i.e.\ $\langle \Phi(0,\bx) \Phi(t,\by) \rangle \simeq 0$. 
The standard approximations 1) and 3) may, in
principle, be evaded. However, if they are not applied then 
a gradient expansion would become too cumbersome to be of use 
in practical calculations. 
\begin{figure}[t]
\begin{center}
\epsfig{file=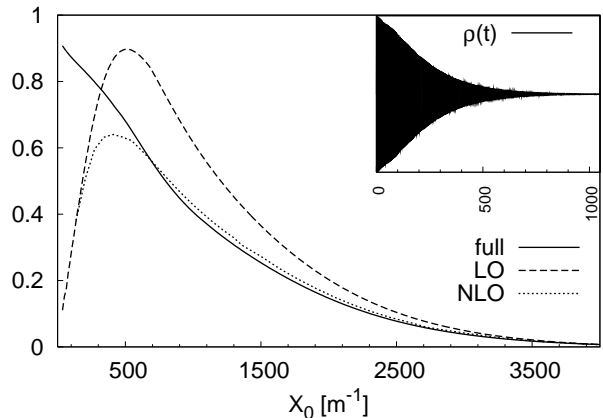,width=8.5cm}
\end{center}
\vspace*{-1.cm}
\caption{
Comparison of LO, NLO and full result for the on-shell $2 \omega 
\partial_{X^0} \tilde{F}$
with $\lambda = 0.5$. The LO transport equation fails to
describe the full results until rather late times. Taking into
account the substantial NLO corrections the gradient expansion
becomes quite accurate for times larger than about $t_{\rm damp}$.
} 
\label{fig:testderiv}
\end{figure}

The NLO transport equations for $\tilde{F}(X;p)= 
\int\! {\rm d}^4 s\, e^{i p s}
F(X+s/2,X-s/2)$ and $\tilde{\varrho}(X;p)= 
-i \int\! {\rm d}^4 s\, e^{i p s}
\rho(X+s/2,X-s/2)$ in Wigner space are given by 
\bea
\lefteqn{\left[ 2 p^{\mu} \partial_{X^{\mu}} 
+ \left( \partial_{X^{\mu}} M^2 \left( X \right) \right) 
\partial_{p_{\mu}} \right] \tilde{F} \left( X, p \right) 
\,=\, \tilde{\Sigma}_{\varrho} 
\left( X, p \right) \tilde{F} \left( X, p \right)  
- \tilde{\Sigma}_F \left( X, p \right) \tilde{\varrho} \left( X, p \right) 
} \qquad\qquad
\nonumber \\
&& +\,  \left\{ \tilde{\Sigma}_F 
\left( X, p \right) , {\rm Re}\tilde{G}_R\left( X, p \right) 
\right\}_{PB} + \left\{ {\rm Re}\tilde{\Sigma}_R 
\left( X, p \right) , \tilde{F} \left( X, p \right) 
\right\}_{PB}, \quad 
\label{eq:NLOF}\\[0.2cm]
\lefteqn{\left[ 2 p^{\mu} \partial_{X^{\mu}} 
+ \left( \partial_{X^{\mu}} M^2 \left( X \right) \right) 
\partial_{p_{\mu}} \right] \tilde{\varrho} \left( X, p \right)} 
\nonumber \\
& = & \left\{ \tilde{\Sigma}_{\varrho} \left( X, p \right) , 
{\rm Re}\tilde{G}_R \left( X, p \right) \right\}_{PB} 
+ \left\{ {\rm Re}\tilde{\Sigma}_R \left( X, p \right) , 
\tilde{\varrho} \left( X, p \right) \right\}_{PB} 
\label{eq:NLOrho} \, ,
\eea
where $G_R(x,y) \equiv \Theta(x^0-y^0)\rho(x,y)$ and 
$\Sigma_R(x,y) \equiv \Theta(x^0-y^0)\Sigma_\rho(x,y)$ 
denote the retarded propagator and self-energy, and $\{ \ldots \}_{PB}$
are the Poisson brackets.
The corresponding LO equations are obtained by neglecting all contributions
of order $\left(\partial_{X^{\mu}} 
\partial_{p_{\mu}}\right)^2$ in (\ref{eq:NLOF}).
The LO equations are typically used to obtain kinetic equations with
the further assumption of a quasi-particle picture. 

Fig.~\ref{fig:testderiv} shows the time evolution of the on-shell 
derivative $2 \omega \partial_{X^0} \tilde{F}$.
The thick curve represents the ``full'' result, which 
is obtained from solving the evolution equations
(\ref{eq:exactevolF}) and (\ref{eq:exactevolrho}) for the
three-loop 2PI effective action.
For comparison, we evaluate the same quantity
using the LO gradient expansion according to Eq.~(\ref{eq:NLOF}). 
For this we evaluate the RHS of the LO terms in 
Eq.~(\ref{eq:NLOF}) using the full result  
for $\tilde{F}$. If the gradient expansion to lowest order 
is correct, then both results have to agree. Indeed, one finds that
the curves for the LO (dashed) and the full result indeed agree at
sufficiently late times. Taking
into account the NLO contributions the agreement can be improved 
substantially. The NLO curve (dotted) in Fig.~\ref{fig:testderiv} 
approaches the full result rather closely, however, they only agree 
{\em after} some characteristic time. The latter is determined
by the time scale $\sim t_{\rm damp}$ for the effective loss of details about
the initial conditions. This can be observed, e.g.,
from the decay of the unequal-time two-point function 
shown in the inset. The latter measures correlations at time $t$ with the
initial state and its decay on the time scale $\sim t_{\rm damp}$ 
coincides rather well with the time for the onset of validity 
of the NLO result. 

We have done a series of comparisons for various couplings in 
Ref.~\cite{Berges:2005ai}
and conclude that for times sufficiently large compared to $t_{\rm damp}$
the gradient expansion seems to converge well. There are sizeable NLO 
corrections already for couplings of about $\lambda \simeq 1/4$. 
Nevertheless, one observes that the NLO result can get rather close 
to the full result for $\lambda \lesssim 1$~\cite{Berges:2005ai}.
Times shorter than about $t_{\rm damp}$ seem clearly to be
beyond the range of validity of transport equations
even for weak couplings. This should make them unsuitable
to discuss aspects of apparent early thermalization and stresses
the need to employ proper equations such as (\ref{eq:exactevolF})
and (\ref{eq:exactevolrho}) for initial-value 
problems.

\section{Beyond 2PI expansions: direct lattice simulations}

We have seen above that the techniques based on two-particle 
or higher irreducible generating functionals are crucial
for our analytical understanding of nonequilibrium quantum fields.
However, analytical approaches necessarily involve 
approximations such as a 2PI loop expansion. 
Nonequilibrium approximations are difficult 
to test for crucial questions of QCD, i.e.\ where strong interactions can
play an important role. Here direct numerical simulations of the quantum field
theory on a space-time lattice, i.e.\ without truncations, could 
boost our knowledge and trigger the development of further approximate 
analytical tools. 

Despite the importance of non-perturbative lattice simulation techniques in 
out-of-equilibrium quantum field theory, these are still in its  
infancies. This is in sharp contrast to well-established thermal equilibrium
methods. 
Equilibrium calculations can typically be based on 
a Euclidean formulation, where the time variable is analytically 
continued to imaginary values. By this the quantum theory is mapped 
onto a statistical mechanics problem, which can be simulated by 
importance sampling techniques. Nonequilibrium problems, 
however, are not amenable to a Euclidean formulation. 
Moreover, for real times standard importance sampling is not possible 
because of a non-positive definite probability measure.
A very interesting recent development employs stochastic 
quantization techniques for real times, which do not require 
a probability distribution~\cite{Berges:2005yt}. 
Clearly, the possibility of direct
simulations in nonequilibrium quantum field theory
would mark a breakthrough not only for the description of 
QCD dynamics. 

In Ref.~\cite{Berges:2005yt} first lattice simulations of nonequilibrium 
quantum fields in Min\-kowski space-time have been presented. 
For the example of a scalar field theory with quartic self-interaction 
this was used to compute the time evolution of
correlation functions and characteristic time scales. 
Starting from a non-thermal initial state, 
the real-time quantum ensemble in 3+1 dimensions 
is constructed by a stochastic process in an additional (5th) 
``Langevin-time'' using the reformulation of stochastic 
quantization for the Minkowskian path 
integral~\cite{stochquant}. In addition to the space-time 
variable $x$ the field $\phi$ depends on the Langevin-time 
parameter~$\vartheta$ with
\bea
\frac{\partial \phi}{\partial \vartheta} 
&=& i \frac{\delta S[\phi]}{\delta \phi}
+ \eta  \, ,
\label{eq:RI}
\eea
where $S$ denotes the classical action and $\eta$ Gaussian or white noise. 
For a real quantum field theory the Langevin dynamics
governs a {\em complex} $\phi = \phi_R + i \phi_I$,  
where the appearance of an imaginary part reflects the
fact that in the quantum theory the field  
picks up a phase by evolving in time. 

The stochastic process (\ref{eq:RI}) 
is associated to a real distribution $P(\phi_R,\phi_I;\vartheta)$ and
averages of observables $A(\phi)$ are given as area integrals 
in the complex field plane:
\beq
\langle A \rangle_\eta = \frac{\int [{\rm d} \phi_R][{\rm d} \phi_I]
A(\phi_R + i \phi_I) P(\phi_R,\phi_I;\vartheta)}{\int [{\rm d} 
\phi_R][{\rm d} \phi_I] P(\phi_R,\phi_I;\vartheta)}
\equiv \frac{\int [{\rm d} \phi_R]
A(\phi_R) P_{\rm eff}(\phi_R;\vartheta)}{\int [{\rm d} 
\phi_R] P_{\rm eff} (\phi_R;\vartheta)}\, , 
\label{eq:obs}
\eeq
where the second equality defines the complex pseudo-distribution 
$P_{\rm eff}(\phi_R;\vartheta)$. The latter is indeed governed
by the analytic continuation of the Fokker-Planck equation
to real times, which admits
the stationary solution~\cite{stochquant}
\bea
\lim_{\vartheta \to \infty} P_{\rm eff}(\phi_R;\vartheta) 
\sim e^{i S[\phi_R] } \, . 
\eea
Thus the approach can in principle be used for a 
Minkowskian theory, with ``ensemble'' averages calculated as averages
along Langevin trajectories. 

Some properties seem to make the approach quite suitable for 
out-of-equilibrium calculations. Firstly,  
nonequilibrium requires specification of an initial state
or density matrix. Therefore, the initial configuration is
fixed which seems to stabilize the procedure. Moreover,
the additional averaging over an
initial density matrix can help to achieve fast convergence.
Secondly, one typically has a good guess for the $3+1$ dimensional starting 
configurations of the Langevin updating procedure:
In contrast to the quantum theory, the corresponding classical 
statistical field theory can be simulated using numerical 
integration and Monte Carlo 
techniques~\cite{Berges:2004yj}.
Using the nonequilibrium classical statistical solution as the 
starting configuration can improve convergence. 
Here the classical field configurations
are obtained by numerically solving the 
classical field equation of motion and sampling over initial conditions,
with nonzero field average and Gaussian fluctuations. It also provides 
a crucial check of the quantum result in some limiting 
cases: For sufficiently large macroscopic field or 
occupation numbers classical dynamics can 
provide a good approximation for the quantum evolution
at not too late times~\cite{Aarts:2001yn}.

\begin{figure}[t]
\vspace{4.9cm} 
\includegraphics{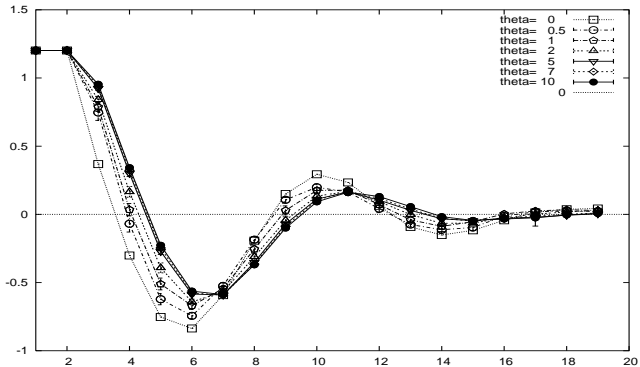} 
\caption{The real part of 
$\langle \int_x \Phi(0,\bx) \Phi(t,\bx) \rangle$ as a funtion of
time $t$ in units of the lattice spacing $a$. As  
starting configuration ($\vartheta =0$)  
the classical result is taken, and the Langevin updating ($\vartheta > 0$)
incorporates quantum corrections~\cite{Berges:2005yt}.} 
\label{fig:int} 
\end{figure}

Fig.~\ref{fig:int} shows the time evolution for 
the connected part of the unequal-time correlator 
${\rm Re} \langle \int_x \Phi(0,\bx) \Phi(t,\bx) \rangle$, which
measures the correlation of the field at time $t$ with the 
initial field. It gives important information about the characteristic 
time scale for the loss of details about the initial conditions,
as explained in Sec.~\ref{intro}.

One finds good convergence properties
of the quantum simulations, which is a remarkable result. 
For given initial field configurations at time $t=0$, very different starting 
configurations for the 3+1 dimensional space-time lattice converge 
to the same nonequilibrium dynamics for all $t > 0$.
To obtain this one had to resolve the
problem of possible unstable dynamics for the updating 
procedure~\cite{Berges:2005yt}.
Though more or less formal proofs 
of equivalence of the stochastic approach and the path integral 
formulation have been given for Minkowski space-time, 
not much is known about the general convergence properties and its 
reliability.  
 
Two procedures can be employed for further tests of the
algorithm. Firstly, one can compare to
analytical approximations based on two-particle 
irreducible effective actions~\cite{simulations}.     
Secondly, going to sufficiently late times 
one can compare to certain thermal equilibrium results from
Euclidean simulations. 

The range of potential applications of first-principles
simulations in nonequilibrium quantum field theory is enormous.
They may be used for out-of-equilibrium as well as 
Minkowskian equilibrium properties extracted at late times.
Possible applications to QCD require implementation in a 
non-Abelian gauge theory, which is work in progress~\cite{simulations}.

We thank Ion-Olimpiu Stamatescu and Christof Wetterich for very fruitful 
collaborations on recent related work.

\vfill\eject
\end{document}